%% file: root.tex
\documentclass[letterpaper, 10 pt, conference]{ieeeconf}  

\IEEEoverridecommandlockouts                              

\usepackage{graphicx}
\usepackage{biblatex}
\usepackage{amsmath,amssymb,amsfonts,dsfont}
\usepackage{stmaryrd}
\usepackage{url}
\usepackage{dsfont}
\usepackage{float}
\usepackage{hyperref}
\usepackage{multirow}
\usepackage{caption}
\usepackage[table,xcdraw]{xcolor}
\include{notations}

\usepackage{adjustbox}
\usepackage{graphicx}
\usepackage{siunitx}
\title{\LARGE \bf
AI vs Humans for the diagnosis of sleep apnea
}

\author{Valentin Thorey$^{1}$*, Albert Bou Hernandez$^{1}$*, Pierrick J. Arnal$^{2}$, Emmanuel H. During$^{3}$
\thanks{*Contributed equally}
\thanks{$^{1}$Algorithms Team, Dreem, Paris, France}%
\thanks{$^{2}$Research Team, Dreem, New York City, USA}%
\thanks{$^{3}$Center for Sleep Sciences and Medicine, Stanford University, Stanford, California, USA}%
}

\begin{document}

©2019 IEEE
\maketitle
\thispagestyle{empty}
\pagestyle{empty}

\begin{abstract}

Polysomnography (PSG) is the gold standard for diagnosing sleep obstructive apnea (OSA). It allows monitoring of breathing events throughout the night. The detection of these events is usually done by trained sleep experts. However, this task is tedious, highly time-consuming and subject to important inter-scorer variability.
In this study, we adapted our state-of-the-art deep learning method for sleep event detection, DOSED, to the detection of sleep breathing events in PSG for the diagnosis of OSA. We used a dataset of 52 PSG recordings with apnea-hypopnea event scoring from 5 trained sleep experts. We assessed the performance of the automatic approach and compared it to the inter-scorer performance for both the diagnosis of OSA severity and, at the microscale, for the detection of single breathing events.
We observed that human sleep experts reached an average accuracy of 75\% while the automatic approach reached 81\% for sleep apnea severity diagnosis. The F1 score for individual event detection was 0.55 for experts and 0.57 for the automatic approach, on average.
These results demonstrate that the automatic approach can perform at a sleep expert level for the diagnosis of OSA.

\end{abstract}


\input{1_introduction}

\input{2_methods}

\input{3_experiments}

\input{4_discussion}

\input{5_conclusion}

\addtolength{\textheight}{-12cm}   



\section*{ACKNOWLEDGMENT}
We thank Dr. Michael E. Ballard, Hugo Jourde, Polina Davidenko, Sarah deLanda and Stephanie Lettieri for their help in realizing the clinical trial at the Stanford Sleep Medicine Center.

\printbibliography[heading=bibintoc] 

\end{document}

%% file: notations.tex
\def \R {\mathbb{R}}

\def \X {\mathcal{X}}

\def \EE {\mathcal{E}}

%% file: 1_introduction.tex
\section{INTRODUCTION}
Obstructive sleep apnea (OSA) is the most common sleep-related breathing disorder, afflicting approximately 25\% of men and 10\% of women \cite{peppard2013increased}. This condition is characterized by repeated episodes of apnea and hypopnea during sleep. Health consequences of OSA range from increased cardiovascular morbidity, motor vehicle accidents due to resulting fatigue, and psychiatric illness to reduced productivity. Despite these consequences, the American Academy of Sleep Medicine estimates that 80\% of people with sleep apnea remain undiagnosed \cite{AASM}. Polysomnography (PSG) is the gold standard method for OSA diagnosis. PSG uses several physiological electrodes to record brain, muscle, and respiratory signals on a patient, usually over a single night in a sleep laboratory. Then, trained sleep experts are able to detect and count breathing events occurring during sleep such as obstructive apneas, central apneas, hypopneas, or mixed apneas. The number of breathing events per hour of sleep is called the apnea-hypopnoea index (AHI). The AHI allows for the evaluation of the severity of OSA, ranging from mild ($5 < AHI < 15$) to severe ($AHI > 30$). However, the process of scoring individual breathing events is tedious, costly, and subject to variability between trained experts \cite{Rosenberg2014}. 

Automatic approaches have been developed to help diagnose sleep apnea. These methods typically extract features from short ($\sim$ 1-second) windows of preprocessed PSG signals. Then, they apply a binary machine learning classifier on each window to detect whether or not a breathing event occured.  For example, in \cite{6153377} the authors preprocess their data with a high-pass filter and a fast Fourier transformation, to extract several window statistics and train an AdaBoost classifier. Alternative models that have been used include Support Vector Machines \cite{khandoker2009support}, K-Nearest Neighbor models \cite{6153377} and shallow Artificial Neural Networks \cite{huang2017novel}.
Other methods advocate for the use of models that can capture the temporal information between successive windows for classification. That is both the case of \cite{song2016obstructive} and \cite{cheng2017recurrent} where discriminative Hidden Markov Model and Long Short Term Memory networks are used, respectively.
All the aforementioned methods rely heavily on handcrafted features and preprocessing, which makes generalization to new patients difficult. Moreover, each PSG signal must be processed in a different way which again increases the difficulty to generalize to new data, devices, and patients.
A last category of methods addresses these limitations by relying on deep learning to extract data-driven relevant features from the windows. In \cite{honore2017large}, a systematic approach for artificial network architecture design called  Progressive  Learning  Network is applied. \cite{choi2018real} proposes a model based on convolutional Neural Networks to leverage their capacity to capture the spatial distribution of the data and \cite{biswal2018expert} goes one step further by combining recurrence with a deep convolutional neural model.
Most methods measure performance on the overall AHI but not on detection of individual breathing events. Moreover, those methods providing individual event detection use fixed-size windows and therefore often cannot capture the correct start and end of individual events. This allows algorithms to provide a diagnosis, but not a detailed analysis of breathing events during the night. Another important drawback of all the aforementioned methods is that they use only one scorer for both training and evaluation. Because of the relatively low inter-scorer agreement, this introduces a positive bias when measuring performance and only gives a partial view of the generalizability of the method.

In this paper, we propose a solution to overcome these limitations. We adapted our state-of-the-art model for sleep microevent detection, DOSED \cite{chambon2018mlsp}, for the detection of breathing events. By doing so, we escape from the fixed window size paradigm and analyze PSG data as sleep experts would. Indeed, the method outputs start and end for every breathing event inside long windows of e.g. $\sim$ 3 minutes. Working on long windows, the approach makes use of the local temporal context to predict events. The model is fully convolutional and highly parallelizable, so it can efficiently process windows of any size. Moreover, by design, the network can be fed with any signal from the PSG relevant for breathing event detection without any special preprocessing.
We evaluated our model in a realistic setup. The model is trained end-to-end using a consensus of apnea-hypopnea events annotations obtained from multiple sleep experts, at different sleep centers. We contextualize the results obtained by comparing them with the inter-scorer agreement. We evaluate performance at the night scale by computing mean AHI error and the Diagnostic Accuracy for the OSA severity. We also compute performance at the microscale by evaluating F1 score on the detection of individual apnea-hypopnea events.

%% file: 2_methods.tex
\section{METHODS}
\paragraph*{Notations}
Let an annotation be a list of centers and durations corresponding to the detected apnea-hypopnea events on a PSG record. An annotation can be made by a sleep expert or by an automatic approach. Let $\X = \R^{C \times T}$ be the set of input windows of signal where $C$ stands for the number of PSG channels considered and $T$ for the number of time steps in a window. For instance, $T = fs \times 180$ where $fs$ is the sampling frequency of the PSG signals for a window of $3$ minutes. We denote $1$ the label associated with apnea-hypopnea events and $0$ the label associated with no events. An event $e \in \EE = \R^2 \times \{0, 1\}$ is defined by a center location time $t^c$, a duration $t^d$ and an event label $l \in \{0, 1\}$.

\paragraph*{Build a consensus annotation from multiple annotations}
We consider annotations made by $n$ sleep experts on a PSG record. To build a consensus from these $n$ annotations, we follow the suggested approach used in \cite{chambon2018dosed}. We generate a binary vector of size $fs \times d$ per scorer from the center and the duration of the events, where $fs$ refers to the sampling frequency of the PSG recording and $d$ to its duration. We then compute the mean vector taking values in $\{0, 1 / n, 2 / n, ..., 1\}$. To retrieve a binary representation, we apply a soft thresholding $\kappa$ on the mean vector. Doing so, we obtain a single consensus that we can encode again with the center and the duration of its events. With $\kappa=1 / n$, the consensus is the union of the events from all the annotations; and with $\kappa=1$, it is the intersection.

\paragraph*{Compute detection performance metrics of an annotation against a consensus annotation}
Given a reference consensus annotation on a PSG record, we want to count the number of true positive (TP), false positive (FP) and false negative (FN) events in a given tested annotation. To do so, the test annotation events are matched with the consensus annotation events based on an overlap criteria: the intersection over union (IoU) \cite{YOLO}. Therefore, we can compute Precision (Pr), Recall (Re), and F1 scores for different minimum IoU thresholds and estimate the exactness of the given tested annotation:

\vspace{-1.5em}
\begin{center}
\begin{equation*}
        Pr = \frac{TP}{TP + FP} \ \ \ \ Re = \frac{TP}{TP + FN}
\end{equation*}
\begin{equation*}
     F1 = 2*\frac{Pr * Re}{Pr + Re}
\end{equation*}
\end{center}

\paragraph*{Automatic Approach}

\begin{figure}[ht!]
\centering
\includegraphics[width=0.9\linewidth]{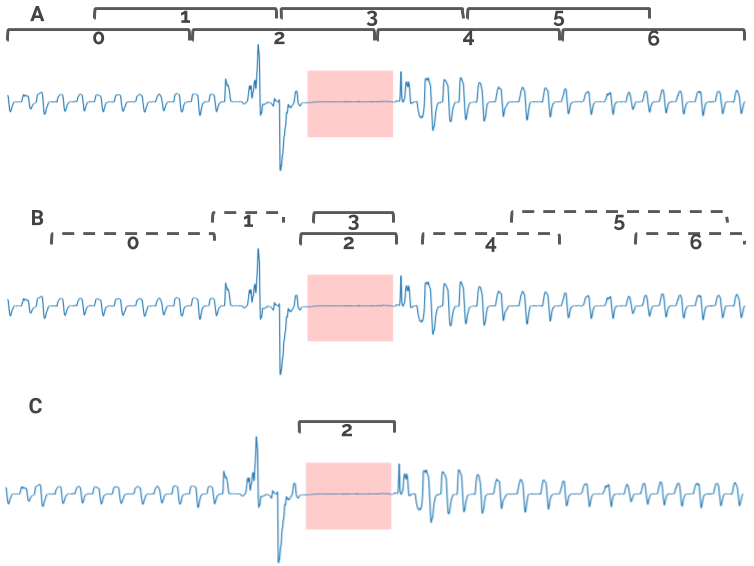}
\caption{\label{fig:scheme}DOSED during prediction. We consider an airflow signal from a PSG on a 3-minute window.  A: 7 default event location of duration 45 seconds with 50\% overlap tiles the 3-minute window. B: the network predicts adjusted location and label for each default event. Dashed events have $l = 0$ (no event) and plain events have $l = 1$ (apnea-hypopnea event) C: $l = 0$ events are removed. Eventually, non-maximum suppression is applied to breathing events to remove overlapping predictions and keep those with high probability.}
\vspace{-0.2em}
\end{figure}

The DOSED approach introduced in \cite{chambon2018mlsp} is used as a guideline for our implementation. We use a convolutional network to predict apnea-hypopnea events on windows of signal from $\X$. The inference procedure works as follows. First, we define $N_d$ default events which are parameterized with a default center time and a default duration. Default events tile each window of signal from $\X$. The convolutional network outputs the probability of an apnea-hypopnea event in each default event location. It also provides localization adjustment to apply to the default event location to fit any apnea-hypopnea event more closely. Then, default events with a probability superior to a cross-validated threshold $\theta$ to contain an apnea-hypopnea event are kept. Eventually, non-maximum suppression (NMS) is applied to remove eventual overlapping predictions of the same event. The figure \ref{fig:scheme} summarizes this inference procedure.

The training procedure follows the one described in \cite{chambon2018mlsp}. For each sample window from $\X$, bipartite matching is used: ground truth events are matched with the default event that overlaps the most with them. Then, remaining default events are matched with the ground truth events if their IoU is above a certain threshold $\gamma$ and are also assigned label $l=1$. Eventually, any remaining default event that does not match any ground truth event is assigned with the label $l=0$.
The model is trained end-to-end by back-propagation by minimizing the multi-objective loss function defined in \cite{chambon2018mlsp}. This loss function combines localization and classification terms for default events with $l=1$. The localization term is ignored for the default events with $l=0$. Moreover, for those events, negative-sampling is used. Only the worst classified among them are considered when computing the classification loss. This intends to alleviate the class imbalance between default events with $l=0$ and those with $l=1$.

The convolutional network architecture has been adapted from \cite{chambon2018dosed} but contains the same building blocks. It is described in detail in Table~\ref{tab:network_architecture}.
First $k$ temporal feature extraction modules are successively applied to raw data from $\X = \R^{C \times T}$ to extract a low dimensional temporal representation $\hat{\X} = \R^{4 * 2^{k} \times T / 2^{k}}$. These feature extraction modules first apply a convolutional layer with zero padding to maintain the temporal dimension. Then batch normalization is applied followed by ReLU activation function. At this step, dropout is applied during training. Eventually, Max pooling is applied to reduce the temporal dimension by a factor 2.
Then, a localization module and a classification module are applied to data from $\hat{\X}$. The localization module outputs a center adjustment and a duration adjustment for each of the $N_d$ default events. This prediction is done with a convolutional layer using $2 \times N_d$ kernels of size $T / 2^{k}$. The classification module outputs a probability of containing a breathing event ($l=1$) or not ($l=0$) for each of the default event. This is obtained by recombining another convolutional layer using $2 \times N_d$ kernels of size $T / 2^{k}$ with a softmax activation.

\input{architecture}

%% file: architecture.tex
\begin{table*}[ht!]
\centering
\vspace{1em}
\begin{adjustbox}{width=500pt}
\renewcommand{\arraystretch}{1.1}
\begin{tabular}{|l|l|l|l|l|l|l|}
\rowcolor[HTML]{FFFFFF} 
                    \hline
                   & layer type                                            & kernel size                                           & kernel \#                                             & output dimension                                      & activation                                            & stride \\ \hline
\rowcolor[HTML]{EFEFEF} 
\renewcommand{\arraystretch}{1}
\begin{tabular}[c]{@{}l@{}}feature extraction\\ k blocks\end{tabular} &  \begin{tabular}[c]{@{}l@{}}Convolution 1D\\ Max Pooling 1D\end{tabular} & \begin{tabular}[c]{@{}l@{}}3\\ 2\end{tabular}           & \begin{tabular}[c]{@{}l@{}}$4 \times 2 ^{k} \ \mbox{if} \ k > 1 \ \mbox{else} \ C$\\ -\end{tabular} & \begin{tabular}[c]{@{}l@{}}$(4 \times 2^{k}, T / 2^{k - 1})$\\ $(4 \times 2^{k}, T / 2^{k})$\end{tabular} & \begin{tabular}[c]{@{}l@{}}relu\\ -\end{tabular}    & \begin{tabular}[c]{@{}l@{}}1\\ 2\end{tabular} \\ \hline
\rowcolor[HTML]{C0C0C0} 
localization       & \begin{tabular}[c]{@{}l@{}}Convolution 1D\\ Reshape\end{tabular}        & \begin{tabular}[c]{@{}l@{}}$T / 2^{k}$\\ -\end{tabular} & \begin{tabular}[c]{@{}l@{}}$2 \times N_{d}$\\ -\end{tabular}                  & \begin{tabular}[c]{@{}l@{}}$(2 \times N_{d}, 1)$\\ $(N_{d}, 2)$\end{tabular}                              & \begin{tabular}[c]{@{}l@{}}-\\ -\end{tabular}       & \begin{tabular}[c]{@{}l@{}}1\\ -\end{tabular}       \\ \hline
\rowcolor[HTML]{C0C0C0} 
classification     & \begin{tabular}[c]{@{}l@{}}Convolution 1D\\ Reshape\end{tabular}        & \begin{tabular}[c]{@{}l@{}}$T / 2^{k}$\\ -\end{tabular} & $2 \times N_{d}$                                                              & \begin{tabular}[c]{@{}l@{}}$(2 \times N_{d}, 1)$\\ $(N_{d}, 2)$\end{tabular}                              & \begin{tabular}[c]{@{}l@{}}-\\ Softmax\end{tabular} & \begin{tabular}[c]{@{}l@{}}1\\ -\end{tabular}       \\ \hline
\end{tabular}
\end{adjustbox}
\caption{Model architecture: First the $k=6$ feature extraction blocks extract a low dimension temporal representation of window sample from $\X$ by successive combination of convolution and max pooling. The localization module outputs localization adjustments for center and duration of the $N_d$ default events. The classification modules outputs the probability of having label $l=0$ and $l=1$ for each of the default $N_d$ events. Localization and classification modules operate in parallel on the output of the feature extraction blocks. Batch dimension is omitted.}
\label{tab:network_architecture}

\vspace{-1.5em}
\end{table*}

%% file: 3_experiments.tex
\section{Experiments}

\subsection{Experimental setup}
\paragraph*{Dataset}
The dataset used in this work was collected at the Stanford Sleep Medicine Center and consists of polysomnography (PSG) recordings from 52 patients (Clinical trial number NCT03657329). Demographics are given in Table \ref{tab:demograpics}.

\input{demographics.tex}

Patients were included in the study based on clinical suspicion for sleep-related breathing disorder. Individuals with a diagnosed sleep disorder different from obstructive sleep apnea were excluded from this study. Individuals suffering from morbid obesity, taking sleep medications, or with complex cardiopulmonary or neurological comorbidities were also excluded. All trial participants gave their informed written consent prior to participation. They received monetary compensation for their time. In total, participants provided 320 hours of sleep data. Recorded PSG signals included the following respiratory signals: chest belt, abdominal belt, SpO2 (oximetry), pressure airflow, nasal airflow, and snoring. The recordings also included EEG channels, and leg muscle and ocular activity. All signals were sampled at 256 Hz.

\paragraph*{Expert annotations}
Sleep breathing events were annotated by 5 different sleep technicians. All scorers were Registered Technologists with at least 5 years of clinical scoring experience across 3 different sleep clinics. Current recommended AASM guidelines were followed. During the process, scorers had access to all recorded signals and scored start time and stop time for each hypopnea, obstructive apnea, central apnea, and mixed apnea event detected. To train the model, annotated events were merged into a single class of apnea-hypopnea event, regardless of their type. Statistics are given in Table \ref{tab:demograpics} and show that patient are essentially suffering from obstructive apnea and hypopnea.

\paragraph*{Automatic approach}
The convolutional network was fed with the $C = 6$ respiratory signal available in the PSG recordings. The sample window considered was 3 minutes long so that $T = 3 * 60 * 256$. This is a typical window size used by sleep experts to annotate breathing events.

To normalize the signal, each channel was first clipped using specific nominal minimum and maximum values. Next, each channel was normalized using min-max normalization to ensure input data were in the specified range $[-0.5, 0.5]$. Finally, the signal was downsampled by a factor of $64$.

There was a high level of variability in the duration of sleep breathing events (10 to 150 seconds, approximately). Therefore, we chose default events sizes of 10, 20, 30, 40, 60, 80, 100, 130 and 150 seconds to tile the 3-minute sample windows. Overlap of 50\% was used for each default event size making the total amount of events $N_d = 92$ default events. $k=6$ feature extraction blocks were used.

This approach was implemented using PyTorch \cite{paszke2017automatic}. Adam optimizer was used with a learning rate of \num{5e-4} and weight decay of \num{e-8}. We trained with minibatch size equal to 128. Maximum IoU overlap for NMS was 0.5. During training, the matching parameter $\gamma$ was set to 0.5. Finally, we used data augmentation techniques, including random noise addition, random signal inversion, and random rescaling of our input data. Dropout was set at 0.1 during training. Windows of signal from $\X$ were drawn at random in training records so that the network could not see the exact same window twice. Each minibatch of data from $\X$ was composed of 50\% windows containing at least one apnea-hypopnea event to alleviate the highly unbalanced number of windows containing events relative to those not containing any events during a night.

To compute the prediction of the automatic approach on the entire dataset, leave-one-out cross-validation was used. The convolutional network was trained 52 times with 31 records for training 20 for validation and 1 for testing for 100 epochs. The training/validation records were selected at random. We used early stopping: at the end of each epoch, the F1 score at $IoU = 0.3$ was computed on the validation set for multiple $\theta$ to select both the best model and best $\theta$ value to use on the final test record. With these settings, convergence was typically achieved after $\sim40-50$ epochs with a F1 score at $IoU = 0.3$ of around $\sim0.6$ on the validation set.

With this setup, inference time was $\sim4$ seconds on an NVIDIA Titan X GPU when predicting breathing events on a full PSG record ($\sim8$ hours). Related code is available at: \url{https://github.com/Dreem-Organization/dosed}.

\subsection{Results}
\begin{figure}[ht!]
\centering
\includegraphics[width=0.99\linewidth]{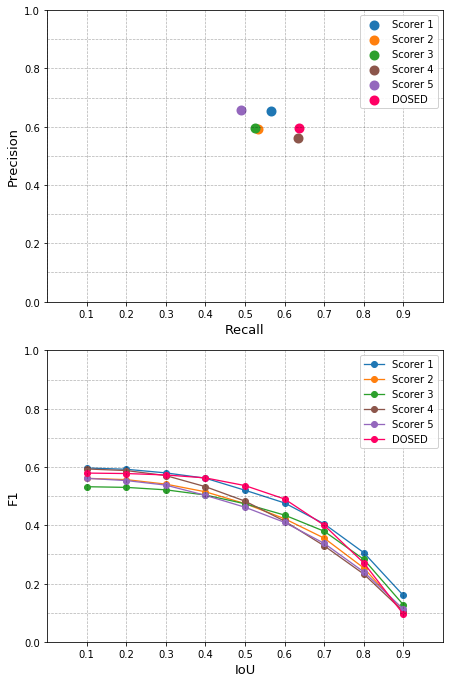}
\caption{\label{fig:results_event}Top: Precision and Recall for each scorer and the automatic approach (DOSED). $IoU = 0.3$. Bottom: Corresponding F1 score as a function of the IoU. Each scorer was compared against the 4 other scorers. DOSED was compared to the 4 best scorers. All the values presented are averaged across the 52 recordings.}
\end{figure}

\paragraph*{Performance at the microscale}
Each scorer annotation was compared against the consensus obtained from the four other scorers. This ensured performance was evaluated in an unbiased manner. The consensus was built with $\kappa = 0.5$. This choice of $\kappa$ retains events tagged by at least 2 of the 4 experts. Metrics were computed per-record and then averaged. To evaluate the performance of the automatic approach, the consensus of the four best scorers was used. The best scorers were selected based on their F1 scores at $IoU = 0.3$. The results are presented for IoU varying between ${0.1, 0.2, ..., 0.9}$ and Precision and Recall are given at $IoU = 0.3$ in Fig. \ref{fig:results_event}. Results show that the automatic approach performs as well as most experts. Expert scoring shows similar level of performance across each sleep expert. However, the expert agreement was relatively low when scoring individual apnea-hypopnea events, consistent with previous results \cite{Rosenberg2014}.

\paragraph*{Performance at the Diagnosis scale}
To evaluate performance at the record scale, two metrics were used: Mean AHI Error and Diagnostic Accuracy of OSA severity. The AHI was computed as the number of breathing events per hour of sleep. The Diagnosis Accuracy was deduced from the AHI based on 3 classes: mild ($AHI \leq 15$), moderate ($AHI \leq 30$) and severe ($AHI > 30$).

Mean AHI Error was computed for each scorer as the absolute distance between the scorer's AHI and the mean AHI of the four other scorers. Diagnostic Accuracy was computed for each scorer by comparison with the consensus AHI, deduced from the mean AHI of the four other scorers. The automatic approach was evaluated in the same way, with a consensus build from the four best scorers. Mean AHI Error and Diagnostic Accuracy were computed for all recordings and then averaged.

Results are summarized in Table  \ref{fig:results_event} and show that the automatic approach performs at a human expert level regarding both the diagnosis of OSA severity and AHI estimation. Diagnostic Accuracy presents a high level of variability for sleep experts because of the edge effects for AHI close to $15$ and $30$.

\input{figures/ahi_results.tex}

%% file: demographics.tex
\begin{table*}
\vspace{1.5em}
\begin{adjustbox}{width=500pt}
\renewcommand{\arraystretch}{1.2}
\begin{tabular}{|c|l|c|c|c|c|c|l|c|c|}
\hline
\rowcolor[HTML]{EFEFEF} 
\#Records & Hours of Sleep  & M/F & Age & BMI & AHI & Obstructive & Hypopnea & Central & Mixed \\ \hline
52 & \multicolumn{1}{c|}{320} & 33/19 & 45.6$\ \pm\ $16.5 & 29.6$\ \pm\ $6.4 & 18.5$\ \pm\ $16.2 & 7.0$\ \pm\ $11 & 7.1$\ \pm\ $6.0 & 0.56$\ \pm\ $1.3 & 0.041 $\ \pm\ $ 0.14 \\ \hline
\end{tabular}
\end{adjustbox}
\renewcommand{\arraystretch}{1}
\caption{Demographics. Breathing events are given by number per hour of sleep on average.}
\label{tab:demograpics}
\vspace{-1.5em}
\end{table*}

%% file: figures/ahi_results.tex

\begin{table}[ht!]
\vspace{0.7em}
\begin{center}
\renewcommand{\arraystretch}{1.3}
\begin{tabular}{
|>{\columncolor[HTML]{EFEFEF}} c|c|c|c|}
\hline
            & \cellcolor[HTML]{EFEFEF}Acc. Diagnosis & \cellcolor[HTML]{EFEFEF}Mean AHI Error & \cellcolor[HTML]{EFEFEF}F1(IoU = 0.3) \\ \hline
scorer 1      & 0.69                             & 4.01$\ \pm\ $3.87                   & 0.58 $ \pm\ $0.23                   \\ \hline
scorer 2       & 0.83                             & 3.82$\ \pm\ $3.93                    & 0.54 $ \pm\ $0.21                \\ \hline
scorer 3      & 0.69                             & 5.00$\ \pm\ $5.72                   & 0.52 $ \pm\ $0.21                           \\ \hline
scorer 4  & 0.73                             & 5.15$\ \pm\ $4.55                   & 0.57 $ \pm\ $0.21                      \\ \hline
scorer 5     & 0.79                             & 4.79$\ \pm\ $4.54                   & 0.54 $ \pm\ $0.21                        \\ \hline
avg. scorers & 0.75              & 4.56$\ \pm\ $4.60                & 0.55$\ \pm\ $0.22              \\ \hline
DOSED       & \textbf{0.81}                   & \textbf{4.69$\ \pm\ $4.25}          & \textbf{ 0.57$\ \pm\ $0.23 } 
\\ \hline
\end{tabular}
\renewcommand{\arraystretch}{1}
\caption{Accuracy Diagnostic for sleep apnea severity, Mean AHI Error and F1 Score at $IoU = 0.3$ for both the five scorers and the automatic approach. Avg. scorers report average global performance across all the 5 scorers. This results shows that DOSED reaches a human performance at both the diagnosis scale and the microscale.}
\label{my-label}
\end{center}
\vspace{-2em}
\end{table}

%% file: 4_discussion.tex
\section{DISCUSSION}
This study suggests that a state-of-the-art deep learning approach for sleep event detection, DOSED, can reach expert human performance when applied to the diagnosis of sleep apnea and detection of breathing events. The novelty in the approach is to mimic the scoring method of a sleep expert by accurately predicting breathing events at the microscale on large time windows, using all the respiratory signals from PSG. This provides not only a reliable diagnostic tool for the evaluation of OSA severity, but also a detailed analysis of breathing events across the entire night.

We analysed our results in comparison to the inter-scorer agreement. The relatively low inter-scorer agreement that was obtained on this study emphasizes the need to rely on multiple expert scorers when developing automatic methods. Using a consensus of multiple scorers provides a better ground truth as a basis for training. Moreover, having multiple scorers  enabled the performance of the model to be evaluated in a naturalistic setting.

Our method could be adapted to predict breathing events by category: hypopnea, obstructive apnea, central apnea, or mixed apnea. This would raise new challenges due to the highly unbalanced nature of each category of events and might require more records to train correctly. This method could also be combined with a sleep stage classifiers to provide detailed information of when breathing events occur to better characterize and diagnose sleep apnea.

%% file: 5_conclusion.tex
\section{CONCLUSIONS}
This paper presents an automatic approach that reaches expert accuracy when diagnosing OSA severity by accurately detecting individual breathing events. The method was designed and evaluated in a realistic setup by comparison with five sleep experts’ annotations and diagnoses. This work shows promising progress towards an automated process for the diagnosis of sleep apnea.